\documentclass[sigconf]{acmart}

\AtBeginDocument{%
  }

\copyrightyear{2024}
\acmYear{2024}
\setcopyright{rightsretained}
\acmConference[L@S '24]{Proceedings of the Eleventh ACM Conference on Learning @ Scale}{July 18--20, 2024}{Atlanta, GA, USA}
\acmBooktitle{Proceedings of the Eleventh ACM Conference on Learning @ Scale (L@S '24), July 18--20, 2024, Atlanta, GA, USA}
\acmDOI{10.1145/3657604.3664688}
\acmISBN{979-8-4007-0633-2/24/07}

\begin{document}

\title{Beyond Repetition: The Role of Varied Questioning and Feedback in Knowledge Generalization}

\author{Gautam Yadav}
\orcid{0000-0002-9247-1920}
\affiliation{%
  \institution{Carnegie Mellon University}
  \streetaddress{5000 Forbes Avenue}
  \city{Pittsburgh}
  \state{PA}
  \country{USA}}
\email{gyadav@andrew.cmu.edu}

\author{Paulo F. Carvalho}
\orcid{0000-0002-0449-3733}
\affiliation{%
  \institution{Carnegie Mellon University}
  \streetaddress{5000 Forbes Avenue}
  \city{Pittsburgh}
  \state{PA}
  \country{USA}}
\email{pcarvalh@cs.cmu.edu}

\author{Elizabeth A. McLaughlin}
\orcid{0000-0003-2650-6504}
\affiliation{%
  \institution{Carnegie Mellon University}
  \streetaddress{5000 Forbes Avenue}
  \city{Pittsburgh}
  \state{PA}
  \country{USA}}
\email{mimim@cs.cmu.edu}

\author{Kenneth R. Koedinger}
\orcid{0000-0002-5850-4768}
\affiliation{%
  \institution{Carnegie Mellon University}
  \streetaddress{5000 Forbes Avenue}
  \city{Pittsburgh}
  \state{PA}
  \country{USA}}
\email{koedinger@cmu.edu}

\renewcommand{\shortauthors}{Yadav et al.}

\begin{abstract}
This study examines the effects of question type and feedback on learning outcomes in a hybrid graduate-level course. By analyzing data from 32 students over 30,198 interactions, we assess the efficacy of unique versus repeated questions and the impact of feedback on student learning. The findings reveal students demonstrate significantly better knowledge generalization when encountering unique questions compared to repeated ones, even though they perform better with repeated opportunities. Moreover, we find that the timing of explanatory feedback is a more robust predictor of learning outcomes than the practice opportunities themselves. These insights suggest that educational practices and technological platforms should prioritize a variety of questions to enhance the learning process. The study also highlights the critical role of feedback; opportunities preceding feedback are less effective in enhancing learning.
\end{abstract}

\begin{CCSXML}
<ccs2012>
   <concept>
       <concept_id>10010405.10010489.10010495</concept_id>
       <concept_desc>Applied computing~E-learning</concept_desc>
       <concept_significance>500</concept_significance>
       </concept>
   <concept>
       <concept_id>10010405.10010489.10010491</concept_id>
       <concept_desc>Applied computing~Interactive learning environments</concept_desc>
       <concept_significance>500</concept_significance>
       </concept>
   <concept>
       <concept_id>10003120.10003121.10011748</concept_id>
       <concept_desc>Human-centered computing~Empirical studies in HCI</concept_desc>
       <concept_significance>300</concept_significance>
       </concept>
 </ccs2012>
\end{CCSXML}

\ccsdesc[500]{Applied computing~E-learning}
\ccsdesc[500]{Applied computing~Interactive learning environments}
\ccsdesc[300]{Human-centered computing~Empirical studies in HCI}

\keywords{Knowledge Generalization, Learning Outcomes, Instructional Design, Repetition, Variability, Feedback, Student Modeling, Model Comparison}


\maketitle

\section{Introduction}
\begin{figure*}[ht]
  \centering
  \includegraphics[width=0.92\linewidth]{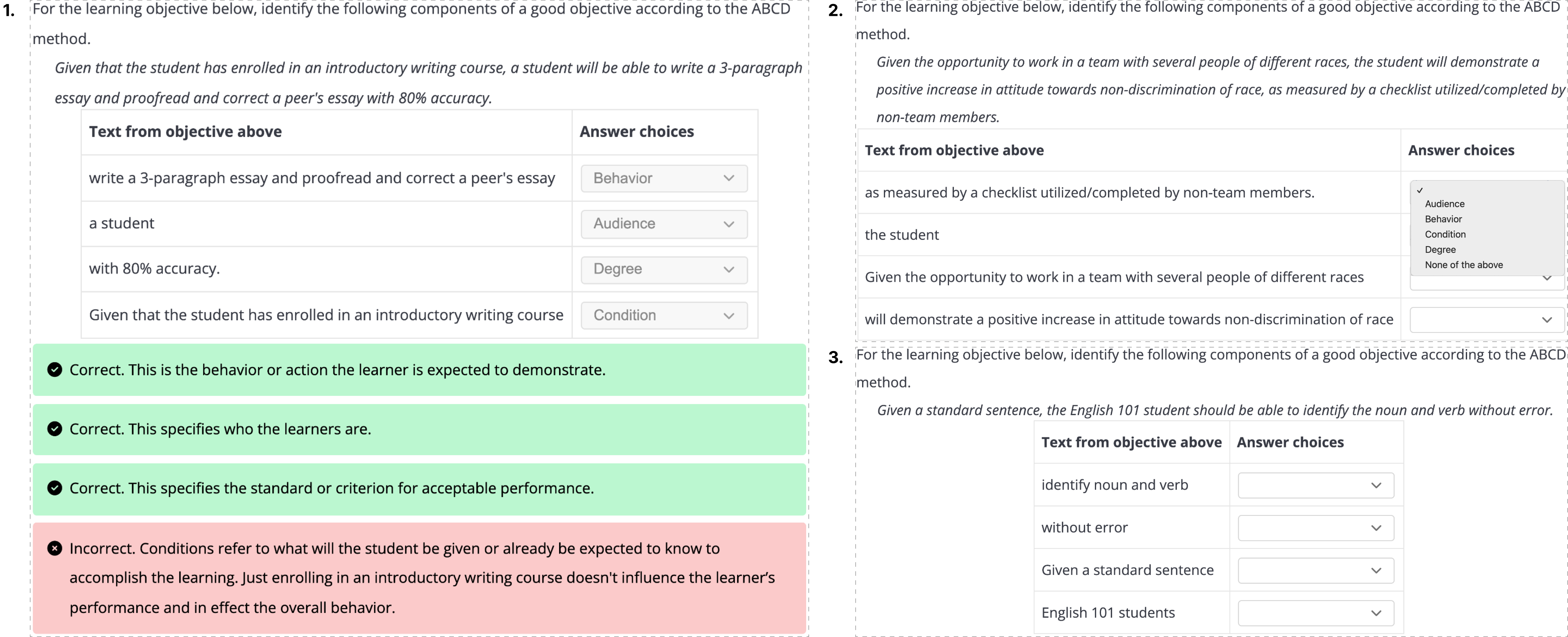}
  \caption{Three unique questions associated with the same knowledge component. Feedback for the first is shown as an example.}
  \label{fig:question_example}
  \Description{
    The image displays an interactive educational interface where users practice identifying components of a good learning objective according to the ABCD method, across three distinct examples.
    The first example focuses on an objective related to writing and proofreading an essay with 80\% accuracy. It includes dropdown menus for selecting correct descriptions for 'Behavior,' 'Audience,' 'Degree,' and 'Condition' based on the given objective.
    The second example presents a learning objective about demonstrating a positive attitude towards non-discrimination of race. It similarly allows users to match components like 'Behavior,' 'Audience,' 'Degree,' and 'Condition' with the appropriate parts of the objective.
    The third scenario involves an objective where students must identify nouns and verbs in a sentence without error. The interface provides dropdown choices for 'Behavior,' 'Degree,' and 'Audience' to be aligned with the correct elements of the sentence.
    Feedback on user selections is immediately provided, highlighting correct answers in green and incorrect in red, thus aiding in understanding how to construct effective learning objectives.
}
\end{figure*}

In online education, immediate feedback with practice questions significantly enhances learning outcomes \cite{lovett2008jime}. A recent study \cite{koedinger2023astonishing} found that students typically require about seven practice opportunities with feedback to master a knowledge component to an 80\% correctness level, emphasizing the importance of practice beyond initial instruction. Despite the benefits, creating effective assessments with feedback is both time-consuming and costly \cite{clifton2010assessing, dibattista2011examination}. One approach to providing more practice involves asking students to repeat assessments once the question pool is exhausted, continuing until they achieve the desired performance for each knowledge component. To illustrate the type of unique questions used in the course, Figure \ref{fig:question_example} presents three unique questions mapped to the same knowledge component, highlighting the variability in question content aimed at enhancing learning generalization.

Firstly, we examine the knowledge component (KC) models to determine whether the timing of explanatory feedback is a more reliable predictor of student learning than practice opportunity. A mapping of KCs to problem steps constitutes a KC Model. Such a model typically utilizes log data to track changes in student performance as the count of opportunities (i.e., interactions or practices with a given KC) increases \cite{stamper2013comparison}. In our implementation, feedback timing differs between inline and quiz assessments: inline provides immediate feedback following the assessment, whereas quizzes provide feedback for all assessment items collectively after the student has submitted the quiz.

After establishing the best-fit model, central to our investigation is the comparative efficacy of unique and repeated questions. Variability in practice opportunities is presumed to hinder immediate performance but enhance learning and transfer \cite{soderstrom2015learning}. This research, conducted within a hybrid graduate-level course, seeks to determine whether the repetition of assessments enhances responses to subsequent, unseen assessments for the same knowledge component. Understanding the best strategies for their development can significantly impact the scalability and effectiveness of online courses.

We also address the frequency of question repetition and its correlation with overall student performance, seeking a threshold where the benefits of repetition decrease, potentially detracting from learning effectiveness.

Our research questions are:

\begin{enumerate}
    \item RQ1: Does learning occur more effectively when feedback is provided after the completion of a quiz, rather than during the quiz as each practice opportunity is attempted?
    \item RQ2: How do unique and repeated opportunities influence students' ability to generalize knowledge to unseen items?
    \item RQ3: How does the frequency of repeated question attempts correlate with students' overall performance?
\end{enumerate}

\section{Related Work}
Previous studies demonstrate the prevalence of a 'doer effect,' where active practice significantly enhances student learning outcomes, surpassing passive learning methods, by comparing model fits from data available in DataShop \cite{koedinger2023astonishing, hou2021drinking, koedinger2015learning}.

The effect of repetition including identical instruction on learning outcomes has classically been studied mainly based on memorization (e.g., word lists) \cite{mathews1973effects, mulligan2013negative}. The role of repetition in learning, especially the contrast between identical task repetition and varying task repetition, has been explored in various contexts but remains underexplored in terms of knowledge generalization. For instance, research in motor skill acquisition in virtual reality settings for children with Developmental Coordination Disorder (DCD) indicated that the type of practice either repetitive versus variable did not significantly affect the transfer of learned skills to real-world tasks \cite{bonney2017learning}. Similarly, studies on language acquisition suggest that while identical task repetition may improve performance in structured tests, it does not significantly outperform task-type repetition in terms of long-term knowledge retention \cite{takimoto2012assessing}.

Knowledge component modeling and learning curve analysis have been used to evaluate student learning in several domains of higher education, such as psychology, biology, statistics, and programming \cite{balter2018estimating, rivers2016learning}. Prior research exploring KC models has focused on course improvement \cite{rivers2016learning} or discovering the best model through the Additive Factors Model \cite{effenberger2020exploration}.

The timing of feedback has also been a critical area of research. Studies examining immediate versus delayed feedback across multiple college classes found no significant differences in learning outcomes between the two at scale \cite{fyfe2021manyclasses}. This finding is particularly relevant to our study as it suggests that if delayed feedback were consistently more effective, the opportunities before feedback would not be as critical for learning.

This study aims to contribute to this ongoing discussion by specifically examining how identical versus different task repetition impacts knowledge generalization. Additionally, we explore alternate KC model mapping based on the timing of feedback, an area that has not been extensively studied.

\section{Methods}
\subsection{Participants and Data Collection}
This study analyzed data from 32 students in a hybrid graduate-level course, "E-Learning Design Principles and Methods" \cite{hou2021drinking} offered through the Open Learning Initiative platform. The course consists of 20 modules with interactive activities, complemented by pre- and post-quizzes. The classroom/remote portion of the course includes lectures, four exams, and two projects. Prior to analysis, we excluded data from students who dropped out of the course. Additionally, opportunities not associated with a specific knowledge component (KC) or labeled with an “unknown” outcome were removed, ensuring a dataset of 30,198 opportunities across 78 KCs.

The interactions with course materials termed "opportunities" include a wide range of activities, from inline formative assessments to review practice quizzes. Student responses to these tasks are automatically tagged as correct when students answer correctly on their first attempt without asking for a hint. Otherwise, the task response is tagged as incorrect. Only the student's first attempt is considered to estimate performance at a given task opportunity, though subsequent student attempts and system feedback are important contributors to learning. We define learning as a positive change in performance and operationalize learning as a reduction in error rate (or increase in correctness rate) over successive opportunities to perform a task associated with a specific knowledge component \cite{koedinger2023astonishing}.

Each module had both a pre-quiz before instruction and a post-quiz, after instruction. Each quiz typically comprised around 8 to 10 selected-response questions drawing from a common pool of questions. Inline activities were specifically designed to mirror the content of quiz and exam questions closely, aligning with specific learning objectives and providing for extra practice. Each quiz had a fixed number of questions for each knowledge component based on their importance as determined by the instructor. Students were permitted to retake post-quiz assessments multiple times, with the highest score across all quiz attempts counted towards their final grade. Students received explanatory feedback on both correct and incorrect responses immediately for inline activities and after quiz completion for each quiz question .

% Figure \ref{fig:question_example} shows an example of three unique questions associated with the same knowledge component. This varied question design is critical for assessing the generalization of knowledge across different situational contexts.

\subsection{Labeling Process}
To precisely measure learning opportunities, we distinguished between opportunities based on each attempt a student made and opportunities adjusted for quiz feedback. We determined the better KC model based on the Akaike Information Criterion (AIC) \cite{stamper2013comparison}.

In the \textbf{Practice Opportunity Labeling} approach, every practice attempt is counted as a separate learning opportunity. This method allows us to track each interaction a student has with the material, classified as either a unique opportunity or a repeat opportunity, based on the sequence of attempts made for a particular KC. Table \ref{tab:freq} provides an example. It illustrates the sequence of learning opportunities a student received for different knowledge components kc1 and kc2 while answering questions (q1-q5) on different quizzes (z1-z3). Focusing on the Practice columns, we see in the second row one prior learning opportunity (Practice Opp=1) and because it was a different question it was a unique opportunity (Unique=1). Skipping to the fourth row (q3), we see 3 prior opportunities (Opp=3) with two that were unique (q1 and q2 in rows 1 and 2) and one that was a repeat (q1 a second time in row 3).

The \textbf{Quiz-Adjusted Opportunity Labeling} columns count opportunities only after a quiz is complete and feedback has been provided on all questions. Thus, we see in row 2, that there are no prior learning opportunities yet (Quiz-Adjusted Opp=0) because the student has yet to receive feedback on their answer to question q1. This adjustment provides insights into how the timing of feedback impacts the learning progression, by only considering those interactions that are reinforced by feedback. For inline questions, the labeling is the same as the Practice Opportunity because immediate feedback is given after each individual assessment.

\begin{table}[ht]
\centering
\caption{Practice Opportunity and Quiz-Adjusted Opportunity Labeling}
\label{tab:freq}
\setlength{\tabcolsep}{2pt}
\begin{tabular}{cccccccccc}
\toprule
\multicolumn{3}{c}{} & \multicolumn{3}{c}{Practice} & \multicolumn{3}{c}{Quiz-Adjusted} \\
\cmidrule(lr){4-6} \cmidrule(lr){7-9}
Question & KC & Quiz & Opp & Unique & Repeat & Opp & Unique & Repeat \\
\midrule
q1 & kc1 & z1 & 0 & 0 & 0 & 0 & 0 & 0 \\
q2 & kc1 & z1 & 1 & 1 & 0 & 0 & 0 & 0 \\
q1 & kc1 & z2 & 2 & 1 & 1 & 2 & 1 & 1 \\
q3 & kc1 & z2 & 3 & 2 & 1 & 2 & 1 & 1 \\
q2 & kc1 & z3 & 4 & 2 & 2 & 4 & 2 & 2 \\
q4 & kc2 & z1 & 0 & 0 & 0 & 0 & 0 & 0 \\
q5 & kc2 & z1 & 1 & 1 & 0 & 0 & 0 & 0 \\
q4 & kc2 & z2 & 2 & 1 & 1 & 2 & 1 & 1 \\
\bottomrule
\end{tabular}
\end{table}

% \begin{table}[ht]
% \centering
% \caption{Example of Practice Opportunity and Quiz-Adjusted Opportunity Labeling: This table illustrates the sequence of learning opportunities for two different knowledge components (KC1 and KC2) across several quizzes (qz1, qz2, qz3) for a particular student. Columns for same-day and different-day repeats, which are based on their step time, are not shown in this table but are part of the overall dataset. The "q1," "q2," etc., represent different questions or attempts within a quiz for a given knowledge component.}
% \label{tab:freq}
% \setlength{\tabcolsep}{2pt}
% \begin{tabular}{cccccccccc}
% \toprule
% \multicolumn{3}{c}{} & \multicolumn{3}{c}{Practice} & \multicolumn{3}{c}{Quiz-Adjusted} \\
% \cmidrule(lr){4-6} \cmidrule(lr){7-9}
% Question & Quiz & KC & Opp & Unique & Repeat & Opp & Unique & Repeat \\
% \midrule
% q1 & qz1 & kc1 & 0 & 0 & 0 & 0 & 0 & 0 \\
% q2 & qz1 & kc1 & 1 & 1 & 0 & 0 & 0 & 0 \\
% q1 & qz2 & kc1 & 2 & 1 & 1 & 2 & 1 & 1 \\
% q2 & qz2 & kc1 & 3 & 1 & 1 & 2 & 1 & 1 \\
% q3 & qz3 & kc1 & 4 & 2 & 0 & 4 & 2 & 0 \\
% q3 & qz3 & kc1 & 5 & 2 & 1 & 4 & 2 & 1 \\
% q1 & qz1 & kc2 & 0 & 0 & 0 & 0 & 0 & 0 \\
% q2 & qz1 & kc2 & 1 & 1 & 0 & 0 & 0 & 0 \\
% q3 & qz1 & kc2 & 2 & 2 & 0 & 0 & 0 & 0 \\
% q2 & qz2 & kc2 & 3 & 2 & 1 & 3 & 2 & 1 \\
% q3 & qz2 & kc2 & 4 & 3 & 0 & 3 & 2 & 0 \\
% q1 & qz2 & kc2 & 5 & 3 & 1 & 3 & 2 & 1 \\
% \bottomrule
% \end{tabular}
% \end{table}

\subsection{Data Analysis}
% We conducted a learning curve analysis to categorize knowledge components (KCs) based on their learning outcomes over successive task opportunities. This analysis was performed using DataShop’s default thresholds, which helped identify distinct learning patterns across KCs by examining changes in error rates.

% We employed a generalized linear mixed model (GLMM) to assess the impact of question repetition and feedback timing on student learning outcomes. The dependent variable in this model is "First.Attempt," which indicates whether a student's first attempt at a question was correct (coded as 1) or incorrect (coded as 0). The model incorporated fixed effects for each type of question repetition—unique, same-day repeated, and different-day repeated—and random effects to account for individual differences in student prior knowledge (intercept) and in KC difficulty (intercept) and learning rate (slope).

For RQ1, we conducted a learning curve analysis to categorize knowledge components (KCs) based on their learning outcomes over successive task opportunities (using DataShop’s default thresholds \cite{datashop_learning_curve}). This method helped us identify distinct learning patterns across KCs by examining changes in error rates. For RQ2, we utilized a generalized linear mixed model (GLMM) to assess the impact of question repetition and feedback timing on student learning outcomes. The dependent variable in this model is 'First.Attempt,' indicating whether a student’s first attempt at a question was correct (coded as 1) or incorrect (coded as 0). This model incorporated fixed effects for each type of question repetition—unique, same-day repeated, and different-day repeated—and random effects to account for individual differences in student prior knowledge (intercept) and in KC difficulty (intercept) and learning rate (slope). For RQ3, we analyzed the frequency of question repetitions and its correlation with overall student performance using correlation and regression analyses. 

% This analytical framework enabled a comprehensive evaluation of how distinct learning opportunities, differentiated by our labeling process, contributed to students' performance and knowledge acquisition.

% Formula: First.Attempt ~ (1 | Anon.Student.Id) + quiz_adj_unique_opp +  
%     quiz_adj_repeat_opp + quiz_adj_different_day_opp + (quiz_adj_unique_opp |  
%     KC..v1.prompt.CTAmultimedia.) + (0 + quiz_adj_repeat_opp |  
%     KC..v1.prompt.CTAmultimedia.) + (0 + quiz_adj_different_day_opp |      KC..v1.prompt.CTAmultimedia.)

\section{Results}
% The learning curve analysis categorized the knowledge components (KCs) based on their learning outcomes into several distinct categories, as shown in Table \ref{tab:kc_category}. This classification helps in understanding the general effectiveness of the instructional strategies employed in the course. The majority of the KCs (88.5\%) fall into the 'Good' category, indicating effective learning outcomes for most of the content covered in the course. The 'Low and Flat' and 'No Learning' categories represent areas where learning outcomes were not optimal, suggesting potential areas for instructional improvement.
% \begin{table}[ht]
% \centering
% \caption{KCs per Category}
% \label{tab:kc_category}
% \begin{tabular}{lcc}
% \toprule
% Category & Number of KCs & \% of total \\
% \midrule
% Low and flat & 1 & 1.3 \\
% No learning & 2 & 2.6 \\
% Still high & 6 & 7.7 \\
% Too little data & 0 & 0 \\
% Good & 69 & 88.5 \\
% \bottomrule
% \end{tabular}
% \end{table}

% Based on the learning curve analysis, we classified the knowledge components (KCs) according to their observed learning outcomes. Using DataShop, with a student threshold set at 10 to filter out less informative data points, we classified KCs into distinct categories:

We classified KCs into distinct categories based on learning curve analysis setting a student threshold at 10 to filter out less informative data points in DataShop:

\begin{itemize}
    \item 'Low and Flat' [1 KC] includes curves where all points are below a 20\% error rate, indicating consistently high performance from the onset.
    \item 'No Learning' [2 KCs] identifies curves with a slope below the 0.001 AFM slope threshold, signifying no significant improvement in student performance over time.
    \item 'Still High' [6 KCs] encompasses curves where the final error rate remains above the 40\% high error threshold, suggesting that students have not reached a satisfactory level of understanding.
    \item 'Good' [69 KCs] denotes curves with a significant positive slope, illustrating that effective learning is occurring as students improve with more opportunities.
\end{itemize}

All KCs met the minimum opportunity threshold of three, so we did not assign any to the 'Too Little Data' category.

\subsection{RQ1: Learning Occurs After Feedback}
The quiz-adjusted model in Table \ref{tab:kc_models}, which only counts opportunities after quiz feedback is provided, fitted the data better (AIC = 35126.2) than the model based on practice opportunity counts (AIC = 35165.4). This improvement indicates that counting opportunities only after quizzes—and thereby incorporating the effect of feedback—results in a more accurate model of student learning.

\begin{table}[ht]
\centering
\caption{Comparison of KC models based on different labeling approaches (practice-based vs quiz-adjusted)}
\label{tab:kc_models}
\setlength{\tabcolsep}{3pt}
\begin{tabular}{lccccc}
\toprule
Labeling & AIC & BIC & logLik & deviance & df.resid \\
\midrule
Practice-Based & 35165.4 & 35248.6 & -17572.7 & 35145.4 & 30188 \\
Quiz-Adjusted & 35126.2 & 35209.3 & -17553.1 & 35106.2 & 30188 \\
\bottomrule
\end{tabular}
\end{table}

\subsection{RQ2: Much Better Generalization from Varied Questions than Repeated Questions}
The generalized linear mixed model (GLMM) analysis, using the Quiz-Adjusted model from Table \ref{tab:kc_models}, was conducted after finding the best-fit model. Table \ref{tab:model_parameters} provides compelling evidence for the superiority of varied questions over repeated ones in fostering student generalization abilities on unseen items. Unique learning opportunities showed a significant positive effect on students’ ability to generalize knowledge to unseen items (p < .001). This result indicates that students benefit considerably more from engaging with new material, supporting the hypothesis that varied questions significantly enhance learning generalization.

For same-day repeated opportunities, the analysis showed a marginally positive effect (p = .064), suggesting that immediate repetition may enhance short-term learning. In contrast, different-day repeated opportunities exhibited a non-significant negative trend (p = .314). This indicates that the benefits of repetition for long-term retention are less clear.

\begin{table}[t]
\centering
\caption{Model parameter estimates}
\label{tab:model_parameters}
\begin{tabular}{lcccc}
\toprule
Parameter & Coef. & Std. Error & z & Pr(>|z|) \\
\midrule
(Intercept) & 0.413 & 0.101 & 4.101 & <.001 \\
unique & 0.092 & 0.013 & 7.020 & <.001 \\
repeat: same day & 0.006 & 0.003 & 1.851 & .064 \\
repeat: different day & -0.018 & 0.018 & -1.006 & .314 \\
\bottomrule
\end{tabular}
\end{table}

\subsection{RQ3: Impact of Repetition}
We initially analyzed the difference in performance between unique and repeated attempts. Our analysis showed that students had an average performance of 64.59\% on unique attempts, whereas on repeated attempts, they scored higher, at 71.31\% (p < .001).

Next, we explored how the frequency of question repetitions correlates with the average quiz scores across the course. We found a significant negative correlation (r = -0.59, p < .001) indicating that students who tend to score lower grades on the quizzes frequently repeat questions. This result is consistent with the course policy that students can retake quizzes when their score is lower than desired such that lower scores yield more repetition.

\section{Discussion and Limitations}
The analysis indicates that unique learning opportunities positively impact performance and, at best, same-day repetition may offer marginal benefits. A key implication is that we should add more questions to our pool of questions so as to maximize practice on unique questions and minimize repetitions of identical questions. Indeed, we have been using generative AI to help us in adding new questions for a given knowledge component.

Another finding is that incrementing the count of learning opportunities per knowledge component \textit{after} a quiz rather than \textit{during} the quiz yields a better prediction of students' future performance. Because students get feedback after the quiz rather than during, this result provides further evidence for the value of feedback in learning --  namely, while there is little or no improvement from question to question without feedback during a quiz, there is improvement revealed after the quiz and receiving feedback.

Our study’s insights come with limitations. Specifically, we combined data from 78 distinct knowledge components, each with varying question pool size and repetition frequency. This amalgamation likely skews results, blending data from frequently repeated questions with those revisited after longer periods. Such an approach may mask the subtleties present when examining individual knowledge components, where repetition's impact on performance could vary significantly. Future research should disaggregate these effects and closely examine the impacts of question pool size and the timing of repetitions within each knowledge component to more precisely determine how these factors influence learning outcomes.

Another limitation is the correlational nature of our findings, as the relationship between unique opportunities and performance may not imply causation. An alternative explanation is that students who engage more with unique opportunities are already stronger learners, while those who repeat questions more frequently might be weaker learners. This explanation seems unlikely for two reasons. First, given our past observation of low variability in student learning rates \cite{koedinger2023astonishing} in online practice with feedback, it is unlikely there are particularly stronger and weaker learners. Second, given our analysis relies on within-student comparisons and controls for student prior knowledge, the result is unlikely based on student differences. Nevertheless, further investigation is warranted.

\section{Conclusion}
% Our findings indicate that engaging with unique learning opportunities substantially boosts performance, more so than engaging in repetitions of the same questions. While repetition initially aids performance, it does not necessarily enhance the ability to generalize to unseen problems, and excessive repetition does not correlate with improved quiz scores. The trend of increased error rates with higher repetition frequency, though not statistically significant, points towards a complex relationship between repetition and learning outcomes.

We presented a novel analysis, incrementing opportunity count within versus after a quiz, that provides further evidence for the importance of feedback during practice in aiding student learning.  More importantly, we found evidence that engaging students with unique learning opportunities is correlated with higher future performance more so than engaging in repetitions of the same questions. While identical question repetition helps performance on the repeated question, it does little to aid generalization to unseen questions. This result is interesting scientifically in that theories of learning that emphasize the role of memory rather than a general concept or skill induction may be interpreted as predicting learning benefits from repeating the same question, whereas we found only a small, non-significant trend for such.  Students experiencing repeated questions appear to memorize the answer verbatim (yielding better performance on repeated questions) but are much less likely to engage in attempts to induce a general concept or skill than when experiencing different questions tapping the same general concept or skill. An important practical implication is the value of having a larger pool of practice questions for each knowledge goal so as to facilitate the benefits of varied practice with unique questions and reduce the chance of repeating identical questions, which may do little more than take student time.

\begin{acks}
We thank Hui Cheng for providing custom reports of the KC Model metrics and her programming expertise.
\end{acks}

\bibliographystyle{ACM-Reference-Format}
\bibliography{PAPER}

\end{document}